\documentclass[twocolumn,superscriptaddress,showpacs,preprintnumbers,amsmath,amssymb,prl]{revtex4}

\usepackage{graphics,epsfig,epsf}
\begin{document}


\title{
Evidence for the exclusive decay         
$B_c^{\pm} \rightarrow J/\psi \pi^{\pm}$ and measurement of the mass of the
$B_c^{\pm}$ meson 
}


\affiliation{Institute of Physics, Academia Sinica, Taipei, Taiwan 11529, Republic of China} 
\affiliation{Argonne National Laboratory, Argonne, Illinois 60439} 
\affiliation{Institut de Fisica d'Altes Energies, Universitat Autonoma de Barcelona, E-08193, Bellaterra (Barcelona), Spain} 
\affiliation{Baylor University, Waco, Texas  76798} 
\affiliation{Istituto Nazionale di Fisica Nucleare, University of Bologna, I-40127 Bologna, Italy} 
\affiliation{Brandeis University, Waltham, Massachusetts 02254} 
\affiliation{University of California, Davis, Davis, California  95616} 
\affiliation{University of California, Los Angeles, Los Angeles, California  90024} 
\affiliation{University of California, San Diego, La Jolla, California  92093} 
\affiliation{University of California, Santa Barbara, Santa Barbara, California 93106} 
\affiliation{Instituto de Fisica de Cantabria, CSIC-University of Cantabria, 39005 Santander, Spain} 
\affiliation{Carnegie Mellon University, Pittsburgh, PA  15213} 
\affiliation{Enrico Fermi Institute, University of Chicago, Chicago, Illinois 60637} 
\affiliation{Joint Institute for Nuclear Research, RU-141980 Dubna, Russia} 
\affiliation{Duke University, Durham, North Carolina  27708} 
\affiliation{Fermi National Accelerator Laboratory, Batavia, Illinois 60510} 
\affiliation{University of Florida, Gainesville, Florida  32611} 
\affiliation{Laboratori Nazionali di Frascati, Istituto Nazionale di Fisica Nucleare, I-00044 Frascati, Italy} 
\affiliation{University of Geneva, CH-1211 Geneva 4, Switzerland} 
\affiliation{Glasgow University, Glasgow G12 8QQ, United Kingdom} 
\affiliation{Harvard University, Cambridge, Massachusetts 02138} 
\affiliation{Division of High Energy Physics, Department of Physics, University of Helsinki and Helsinki Institute of Physics, FIN-00014, Helsinki, Finland} 
\affiliation{University of Illinois, Urbana, Illinois 61801} 
\affiliation{The Johns Hopkins University, Baltimore, Maryland 21218} 
\affiliation{Institut f\"{u}r Experimentelle Kernphysik, Universit\"{a}t Karlsruhe, 76128 Karlsruhe, Germany} 
\affiliation{High Energy Accelerator Research Organization (KEK), Tsukuba, Ibaraki 305, Japan} 
\affiliation{Center for High Energy Physics: Kyungpook National University, Taegu 702-701; Seoul National University, Seoul 151-742; and SungKyunKwan University, Suwon 440-746; Korea} 
\affiliation{Ernest Orlando Lawrence Berkeley National Laboratory, Berkeley, California 94720} 
\affiliation{University of Liverpool, Liverpool L69 7ZE, United Kingdom} 
\affiliation{University College London, London WC1E 6BT, United Kingdom} 
\affiliation{Massachusetts Institute of Technology, Cambridge, Massachusetts  02139} 
\affiliation{Institute of Particle Physics: McGill University, Montr\'{e}al, Canada H3A~2T8; and University of Toronto, Toronto, Canada M5S~1A7} 
\affiliation{University of Michigan, Ann Arbor, Michigan 48109} 
\affiliation{Michigan State University, East Lansing, Michigan  48824} 
\affiliation{Institution for Theoretical and Experimental Physics, ITEP, Moscow 117259, Russia} 
\affiliation{University of New Mexico, Albuquerque, New Mexico 87131} 
\affiliation{Northwestern University, Evanston, Illinois  60208} 
\affiliation{The Ohio State University, Columbus, Ohio  43210} 
\affiliation{Okayama University, Okayama 700-8530, Japan} 
\affiliation{Osaka City University, Osaka 588, Japan} 
\affiliation{University of Oxford, Oxford OX1 3RH, United Kingdom} 
\affiliation{University of Padova, Istituto Nazionale di Fisica Nucleare, Sezione di Padova-Trento, I-35131 Padova, Italy} 
\affiliation{University of Pennsylvania, Philadelphia, Pennsylvania 19104} 
\affiliation{Istituto Nazionale di Fisica Nucleare Pisa, Universities of Pisa, Siena and Scuola Normale Superiore, I-56127 Pisa, Italy} 
\affiliation{University of Pittsburgh, Pittsburgh, Pennsylvania 15260} 
\affiliation{Purdue University, West Lafayette, Indiana 47907} 
\affiliation{University of Rochester, Rochester, New York 14627} 
\affiliation{The Rockefeller University, New York, New York 10021} 
\affiliation{Istituto Nazionale di Fisica Nucleare, Sezione di Roma 1, University of Rome ``La Sapienza," I-00185 Roma, Italy} 
\affiliation{Rutgers University, Piscataway, New Jersey 08855} 
\affiliation{Texas A\&M University, College Station, Texas 77843} 
\affiliation{Istituto Nazionale di Fisica Nucleare, University of Trieste/\ Udine, Italy} 
\affiliation{University of Tsukuba, Tsukuba, Ibaraki 305, Japan} 
\affiliation{Tufts University, Medford, Massachusetts 02155} 
\affiliation{Waseda University, Tokyo 169, Japan} 
\affiliation{Wayne State University, Detroit, Michigan  48201} 
\affiliation{University of Wisconsin, Madison, Wisconsin 53706} 
\affiliation{Yale University, New Haven, Connecticut 06520} 
\author{A.~Abulencia}
\affiliation{University of Illinois, Urbana, Illinois 61801}
\author{D.~Acosta}
\affiliation{University of Florida, Gainesville, Florida  32611}
\author{J.~Adelman}
\affiliation{Enrico Fermi Institute, University of Chicago, Chicago, Illinois 60637}
\author{T.~Affolder}
\affiliation{University of California, Santa Barbara, Santa Barbara, California 93106}
\author{T.~Akimoto}
\affiliation{University of Tsukuba, Tsukuba, Ibaraki 305, Japan}
\author{M.G.~Albrow}
\affiliation{Fermi National Accelerator Laboratory, Batavia, Illinois 60510}
\author{D.~Ambrose}
\affiliation{Fermi National Accelerator Laboratory, Batavia, Illinois 60510}
\author{S.~Amerio}
\affiliation{University of Padova, Istituto Nazionale di Fisica Nucleare, Sezione di Padova-Trento, I-35131 Padova, Italy}
\author{D.~Amidei}
\affiliation{University of Michigan, Ann Arbor, Michigan 48109}
\author{A.~Anastassov}
\affiliation{Rutgers University, Piscataway, New Jersey 08855}
% ...
\collaboration{CDF Collaboration}
\noaffiliation

\date{\today}

\begin{abstract}
We report first evidence for a fully reconstructed decay mode of  
the $B_c^{\pm}$ meson in the channel $B_c^{\pm} \rightarrow J/\psi \pi^{\pm}$,
with $J/\psi \rightarrow \mu^+\mu^-$. The analysis is based on an 
integrated luminosity of 360 pb$^{-1}$ in $p\bar{p}$ collisions at 
1.96 TeV center of mass energy collected by 
the Collider Detector at Fermilab. We observe 14.6 $\pm$ 4.6 signal events 
with a
background of 7.1 $\pm$ 0.9 events, and  
a fit to the $J/\psi \pi^{\pm}$ mass spectrum yields
a $B_c^{\pm}$ mass of  
6285.7 $\pm$ 5.3(stat) $\pm$ 1.2(syst) MeV/$c^2$.
The probability of a peak of this magnitude occurring by random
fluctuation in the search region is estimated as 0.012\%. 

\end{abstract}


\maketitle

 Within the standard model of elementary particles, five of the six
different kinds of quarks combine in quark-antiquark pairs to
make mesons.
The $B_c^{\pm}$ meson combines the two heaviest of these
quarks as a bottom-charm quark-antiquark pair.  
Although it has been observed in semileptonic decay 
modes \cite{CDFRunI_bc, D0RunII_bc}, up to now 
no evidence for the $B_c^{\pm}$ has been found in fully reconstructed decay 
modes \cite{CDFI, DELPHI, ALEPH, OPAL}.
Consequently, its mass $M(B_c)$ has not been measured with good
precision.

Nonrelativistic potential models predict the $\bar{b}$ and $c$ quarks
to be tightly bound with a ground state mass in the approximate
range 6200-6300 MeV/$c^2$ \cite{Rosner_91, Quigg_94,
Godfrey_04}. Recent QCD-based perturbative computations up 
to  ${\cal
O}(\alpha_s^4)$ 
predict $M(B_c)$ to be 6307 $\pm$ 17 MeV/$c^2$
\cite{Brambilla_2002, QWGYR}. Most recently,
a three-flavor lattice QCD calculation 
obtains 
$M(B_c) =$ 6304 $\pm$ 12(stat $\oplus$ syst) $^{+18}_{-0}$(cutoff effects)~MeV/$c^2$ \cite{lattice_2004}. 

Several of the predicted $B_c^{\pm}$ decay modes contain a $J/\psi$
meson \cite{Kiselev_03}. These are among the most easily reconstructable
$B_c^{\pm}$ decays at CDF, owing to an efficient dimuon trigger giving high purity 
$J/\psi \rightarrow \mu^+\mu^-$ reconstruction.  The CDF
collaboration made the first observation of the $B_c^{\pm}$ meson in the
semileptonic decay channels $B_c^{\pm} \rightarrow J/\psi l^{\pm} \nu_l X$,
in a sample of 110 pb$^{-1}$ of data at $\sqrt{s} = 1.8$ TeV in Run I
at the Tevatron \cite{CDFRunI_bc}. 
The symbol $X$ denotes possible undetected decay particles.
With a signal of
20.4$^{+6.2}_{-5.5}$ events, the $B_c^{\pm}$ mass was measured to be 6.40
$\pm$ 0.39(stat) $\pm$ 0.13(syst) GeV/$c^2$.  Recently, the D0
Collaboration reported a preliminary observation of a $B_c^{\pm}$ signal in
the decay channel $B_c^{\pm} \rightarrow J/\psi \mu^{\pm} \nu_{\mu} X$ in 
a sample
of 210 pb$^{-1}$ of Run II data \cite{D0RunII_bc}.

In this paper we report first evidence for the $B_c^{\pm}$ meson in the fully
reconstructed decay channel $B_c^{\pm} \rightarrow J/\psi \pi^{\pm}$,
with $J/\psi \rightarrow \mu^+\mu^-$. The analysis is based on a
dataset of 360 pb$^{-1}$ in $p\bar{p}$ collisions collected at
$\sqrt{s}=1.96$ TeV by CDF at the Tevatron during Run II.

The CDF II detector consists of a magnetic spectrometer surrounded by
calorimeters and muon chambers and is described in detail
elsewhere~\cite{cdfIIdet}. The components most relevant to this analysis
are briefly described here.
The tracking system is in a 1.4 T axial magnetic field and consists of
a silicon microstrip 
detector (L00, SVX, ISL, in increasing order of radius) \cite{cdfSVXII, cdfISL,
cdfL00} surrounded by an open-cell wire drift chamber (COT)
\cite{cdfCOT}.
The muon detectors used for this analysis are the central muon drift
chambers (CMU), covering the pseudorapidity range $|\eta|<0.6$
\cite{Ascoli, Dorigo}, and the extension muon drift chambers (CMX), covering
$0.6<|\eta|<1.0$.  Cylindrical coordinates are used with the $+$z axis in the proton 
beam direction.

This measurement uses events containing pairs of muons, each with $|\eta| <$
1.0, selected with a three-level trigger.  At the first trigger level,
muon-candidate track segments in CMU and CMX are matched to COT tracks
obtained with a hardware processor \cite{Thomson}. Dimuon triggers use
combinations of CMU-CMU and CMU-CMX muons with $p_T> $1.5 (2.0)
GeV/$c$ for CMU (CMX) muons,
where $p_T$ is the momentum transverse to the beamline.  
At the second level, opening angle
and opposite-charge cuts are imposed on the muon pairs. At the third
level, three dimensional (3-D) tracking is performed to select
muon pairs with invariant mass, $M(\mu^+\mu^-)$, between 2700 and 4000
MeV/$c^2$. 
  
To reconstruct the $B_c^{\pm} \rightarrow J/\psi \pi^{\pm}$ decay
offline, we make several requirements on the quality of the tracks and
the $J/\psi$ candidate.  To ensure good primary and secondary vertex 
resolution, 
each track must have an $r-\phi$ position measurement on at least three of
five SVX layers.  For $J/\psi$ identification, we require 
matching between the COT muon tracks and the muon chamber track
segments. In addition, we require that 
$3042 < M(\mu^+\mu^-) < 3152$ MeV/$c^2$,
the average $J/\psi$ mass resolution in our sample being 14 MeV/$c^2$.
Each other charged particle track with $p_T>400$ MeV/$c$ is treated
as a pion candidate to be combined with the $J/\psi$. The pion candidate and the
two muons are then fitted to a common 3-D vertex, with $M(\mu^+\mu^-)$
constrained to the world average $J/\psi$ mass value \cite{PDG2004}.
All combinations for which the vertex fit converged are retained.
The primary
vertex position is calculated from the other tracks in each event.

The $B_c^{\pm}$ search was performed using 
the following analysis method.
The mass values of the $J/\psi \pi^{\pm}$ combinations in the search
window 5600 $< M(J/\psi \pi^{\pm}) <$ 7200 MeV/$c^2$,
referred to as $B_c^{\pm}$ candidates, were temporarily hidden.
The search window was chosen to correspond to the $\pm$2 standard
deviation region around the CDF Run~I measurement of the $B_c^{\pm}$ mass
\cite{CDFRunI_bc}; it is
approximately 100 times wider than the expected $B_c^{\pm}$ mass resolution. 
 
In order to optimize the significance of a possible $B_c^{\pm}$ signal, 
we varied the selection criteria to maximize
the function 
$Q =
S_F/(1.5 + \sqrt{B_{av}})$ \cite{Punzi}.  
Here, $S_F$ is the accepted fraction of signal events, in this case
taken from a Monte Carlo sample, and  the background $B_{av}$ is the number of 
selected $B_c^{\pm}$
candidates within the search window, scaled to correspond to a mass range
of 63 MeV/$c^2$, based on the average mass resolution of a $B_c^{\pm}$
candidate within the search window.  The term 1.5 is appropriate for
optimizing a search for a signal at least $3\sigma$ above background
fluctuations. The distributions of the selection variables for the
signal events were evaluated using samples of simulated
$B_c^{\pm}\rightarrow J/\psi \pi^{\pm}$ decays. These were generated
with a $B_c^{\pm}$ mass of 6400 MeV/$c^2$, a lifetime of 0.46 ps
\cite{CDFRunI_bc}, and $p_T$ and rapidity distributions according to
a leading order perturbative QCD calculation~\cite{Chang}. A harder
$p_T$ spectrum \cite{Bere} was used as an alternative to check the
stability of the optimal selection criteria; these were not very
sensitive to variations of the $p_T$ spectrum or the assumed lifetime
within its experimental uncertainty. The Monte Carlo $B_c^{\pm}$ decays
were processed with full detector simulation and the same
trigger and reconstruction criteria as the data.
The distributions of the selection variables for the background were
taken from the data in the search window, in which the 
contribution from a signal is expected to be small.  

Optimized cuts were determined for the following selection variables: the
$J/\psi \pi^{\pm}$ three-track 3-D vertex
fit ($\chi^2 <$ 9 for four degrees of freedom), 
the pion track contribution to
the vertex fit ($\chi_{\pi}^2 <$ 2.6), the impact parameter in $r-\phi$ of the
$B_c^{\pm}$ candidate with respect to the primary vertex 
($<$ 65 $\mu$m), the maximum $ct$ where $t$ is the proper decay time of the 
$B_c^{\pm}$ candidate
 ($<$ 750 $\mu$m), the transverse momentum of
the pion ($>$ 1.8 GeV/$c$), the 3-D angle between the
momentum of the $B_c^{\pm}$ candidate and the vector joining the
primary to the secondary vertex ($\beta <$ 0.4 rad), and the
significance of the projected decay length of the $B_c^{\pm}$ candidate onto its 
transverse momentum direction ($L_{xy}/\sigma(L_{xy}) >$ 4.4).
  After these selections, 390 
candidates remain in the search window, with no two candidates from the same beam crossing.

\begin{figure}
\footnotesize
\begin{center}
        \mbox{
\epsfysize=2.5in
\epsffile {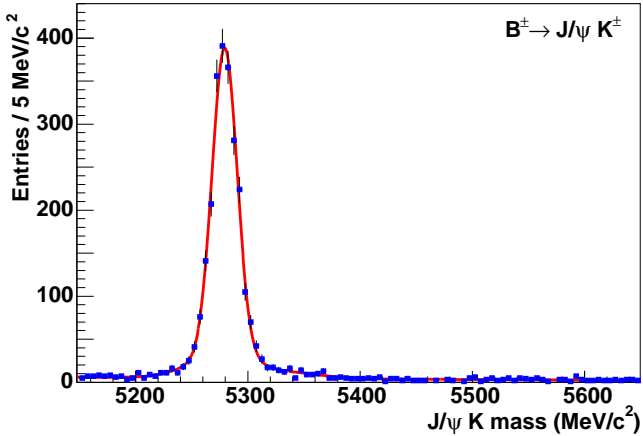}
}
\vspace*{-0.10in}
\caption{ 
The invariant mass distribution of the 
$B^{\pm} \rightarrow J/\psi K^{\pm}$ candidates. The curve is a fit to the
data.
}
\end{center}
\label{upsxsec}
\end{figure}

A sample of $B^{\pm}$ mesons, reconstructed in the decay mode $B^{\pm}
\rightarrow J/\psi K^{\pm}$, was analyzed as a control sample in order
to check our understanding of the reconstruction of the relevant
variables in the simulation.
The $B^{\pm}~ \rightarrow~ J/\psi~ K^{\pm}$ decay topology is the same as that
of $B_c^{\pm}~ \rightarrow~ J/\psi~ \pi^{\pm}$, apart from the different
masses and lifetimes.
The $B^{\pm}$ mass distribution, shown
in Fig. 1, was obtained using the same selection requirements as optimized for
the $B_c^{\pm}$ candidates, but without the cut on maximum $ct$. 
A total of 2378 $\pm$ 57 $B^{\pm} \rightarrow J/\psi K^{\pm}$ signal
events is found, with a fitted mass of 5279.0 $\pm$ 0.3 MeV/$c^2$.
The fit takes into account a small contribution from the Cabibbo-suppressed
decay  $B^{\pm} \rightarrow J/\psi \pi^{\pm}$. 
The average mass resolution is 11.5 $\pm$ 0.3 MeV/$c^2$, in agreement with
the simulation, which can
thus be used with confidence to evaluate the expected mass resolution
for $B_c^{\pm}$ decays.
The $B^{\pm}$ yield is used to calculate the expected $B_c^{\pm}$ yield. The 
relative trigger and reconstruction efficiency, 
$\epsilon_{B_c^{\pm}}$/$\epsilon_{B^{\pm}}$, 
is in the range 35\%-85\%, with uncertainties arising from the $B_c^{\pm}$
$p_T$ spectrum and the $B_c^{\pm}$ lifetime. On the basis of the
$B^{\pm}$ yield, previous CDF cross section measurements
\cite{CDFRunI_bc}, and theoretical calculations
\cite{Kiselev_03,ChangChen, Berez, Anisimov, Colangelo, Hady, yellow} 
of the branching fractions of the 
$B_c^{\pm} \rightarrow J/\psi \pi^{\pm}$ and $B_c^{\pm} \rightarrow J/\psi l^{\pm} \nu$
decay modes, a $B_c^{\pm}$ yield in the range of 10 to 50 events is expected. 

A search procedure was then defined to identify any possible
signal in the data and to estimate its significance.
This was based on a
scan of the search region in 10 MeV/$c^2$ intervals, with a sliding
fit window extending from $-$100 MeV/$c^2$ to $+$200 MeV/$c^2$ in mass
around each nominal peak position, $m$.  This window was chosen to
minimize possible contributions from partially reconstructed $B_c^{\pm}$
decays below the peak position (e.g.\ into $J/\psi$ and more than one 
additional particle). 
For each value of $m$, a fit function was defined as a Gaussian signal
with mean $m$, combined with a linear background term. The
Gaussian resolution was fixed as a linear function of $m$ based on Monte Carlo
simulation, and varied from 13 to 19 MeV/$c^2$ over the search
region.  The three fit parameters were the number of signal (S) and 
background (B) events and the linear background slope. The
output of a scan was defined to be the largest value of
$\Sigma = S/(1.5 + \sqrt{B})$,
$\Sigma_{max}$, obtained from the 131 fits performed in the mass
interval 5700 $\leq M(J/\psi \pi^{\pm}) \leq$ 7000 MeV/$c^2$.

The distribution of $\Sigma_{max}$ for the null hypothesis was obtained from
Monte Carlo experiments \cite{Rolke}, in which the mass spectra were derived 
from a smooth background model.  This model was necessarily approximate 
owing to the initially hidden mass distribution.
The model consisted of a linear background,
to describe combinatoric events, 
 and a ``physical'' background to describe partially reconstructed
$B_c^{\pm}$ decays
in the mass range below 6400 MeV/c$^2$. 
Studies showed that the main source of combinatoric
background are events in which a genuine J/psi is paired with an uncorrelated track.
The shape of the physical
background was based on Monte Carlo simulations of inclusive 
$B_c^{\pm} \rightarrow J/\psi X$ decays, with branching ratios 
taken from Ref.~\cite{Kiselev_03}.

Applied to the 390 event data sample, the scan procedure
found a $\Sigma_{max}$ near $m=$ 6290 MeV/c$^2$, which is compatible 
with a $B_c^{\pm}$ signal of
19 $\pm$ 6 events.
Using a large set of Monte Carlo simulations we modeled the shape of the 
observed background, and, analysing it in the same way as
the data, evaluated the probability that a random enhancement
has a $\Sigma_{max}$ value exceeding that of the data.  This probability
was found to be 0.17\%. 

After the above steps had been performed, further 
checks on the previously hidden events revealed 
that the existing pion selection allowed two classes of fitted tracks
that were unsuitable for the $B_c^{\pm}$ search.
The first class had insufficient number of COT hits to give 
good mass resolution and so 
was not compatible with a search for a narrow Gaussian signal;
the second class had poor SVX resolution in the $z$ 
direction and was dominated by combinatorial
background. The above two classes of events contributed 10\% to the $B^{\pm}$ 
signal; they would be expected to contribute
fewer than two events to the $B_c^{\pm}$ signal, but they
increase the combinatorial background by about 40\% over 
the $J/\psi \pi^{\pm}$ mass range.  After removal of both
classes of poor quality tracks in addition to the original optimized cut 
selection, 220 candidates remained. 
These were required to have good
SVX $z$ resolution on both the pion track and at
least one of the muon tracks.  This final track selection, which maximizes
Q, 
is therefore not fully blind; it is based also on the observed properties 
of the $B^{\pm}$ signal and the overall properties of the 
$B_c^{\pm}$ candidate sample.

Figure 2 shows the mass spectrum for the 220 event sample.
The main features are the $B_c^{\pm} \rightarrow J/\psi \pi^{\pm}$ signal 
peak near 6290 MeV/c$^2$,
a linear combinatorial background above this peak, and a broad
enhancement below the peak which can be attributed to
the physical background from partially reconstructed $B_c^{\pm}$
decays. 
We perform a global unbinned likelihood fit over the
entire mass range to obtain the mass and yield for the $B_c^{\pm}$ 
signal.
The fit included a Gaussian signal with a variable mass
but with a resolution whose mass-dependent value was determined by 
the Monte Carlo simulation,
together with background modeled as a linear combinatorial term
and a broad low-mass Gaussian contribution for the physical background.
A signal of 14.6 $\pm$ 4.6
events is obtained centered at a mass of 6285.7 $\pm$ 5.3 MeV/$c^2$.
The standard deviation of the Gaussian at the central value of the signal mass 
is 15.5 MeV/$c^2$. The background within a
region of $\pm$2 standard deviations from this mass value is
7.1 $\pm$ 0.9 events. The statistical significance
of the signal is discussed below. 
Within the signal region, the distributions of the selection variables
agree within statistics with those of the Monte Carlo simulation.

Systematic uncertainties on the $B_c^{\pm}$ mass determination due to
measurement uncertainties on the track parameters ($\pm0.3$ MeV/$c^2$)
and the momentum scale ($\pm0.6$ MeV/$c^2$) are evaluated from the
corresponding uncertainties on the $B^{\pm}$ mass analysis
\cite{systematics}.  Further uncertainties are due to the possible
differences in the $p_T$ spectra of the $B^{\pm}$ and $B_c^{\pm}$ mesons
($\pm$0.5 MeV/$c^2$) and our limited knowledge of the background shape
used in the final mass fit as well as uncertainty in the signal 
width ($\pm$0.9 MeV/$c^2$) \cite{nicolas}.  The total systematic
uncertainty is evaluated to be $\pm1.2$ MeV/$c^2$.

The signal peak is robust under variations of the pion track quality
selection. 
We have investigated several methods for determining the best
figure of significance for such a peak over a broad mass range.
The method that gives the best sensitivity to a real signal is based on
the standard significance measure $S/\sqrt{B}$.
We repeated the Monte Carlo scans for the new track selection to
determine the null hypothesis distribution for $S/\sqrt{B}$. 
Applying to the 
Monte Carlo simulations the same global fit method as to the data,
 we find that the probability that a random enhancement anywhere in the 
range 5800-7000 MeV/$c^2$ exceeds the value of $S/\sqrt{B}$ for the experimental peak is 0.012\%.

\begin{figure}
\hfil
\epsfysize=2.5in
\epsffile {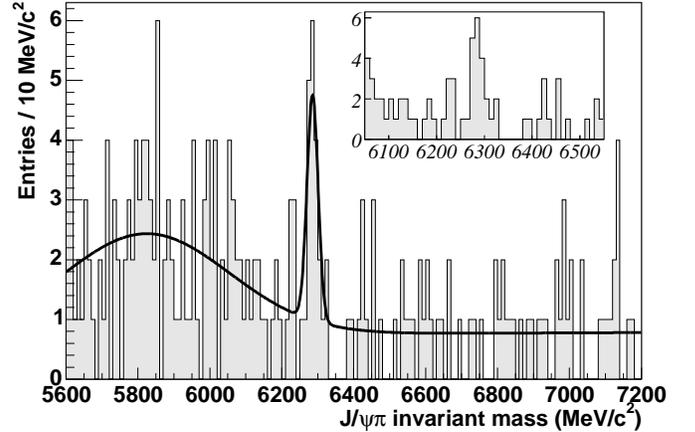}
\hfil
\hfil
\label{fig:mass}
\vspace*{-0.10in}
\caption{ 
The invariant mass distribution of the $J/\psi \pi^{\pm}$ candidates and 
results
of an unbinned likelihood fit in the search window. The inset shows the peak 
section of the distribution.
The broad enhancement below 6.2 GeV/c$^2$ is attributable to partially
reconstructed $B_c^{\pm}$ mesons. 
}
\end{figure}

In view of the limited statistics of the observed mass peak, 
an independent consistency check was performed.  If the
mass peak is due to fully reconstructed 
$B_c^{\pm}\rightarrow J/\psi\pi^{\pm}$
decays, partially reconstructed $B_c^{\pm}\rightarrow J/\psi+$track$+X$  
decays should be detectable in the mass region
below the peak but not in the region above.  The
pion candidate in partially reconstructed decays should have a small
impact parameter $d_{xy}$ relative to the $J/\psi$ vertex, consistent
with being physically associated with it, whereas the pion candidate
in combinatorial background events should have a broad $d_{xy}$
distribution reflecting random association with the $J/\psi$ vertex.
 
To investigate this, we relax the cuts on $\beta$, 
the impact
parameter of the $B_c^{\pm}$ candidate, and the $\chi^2$ of the 
 3-D vertex fit, 
so as to make 
a signal in the $d_{xy}$ distribution visible over the
broader combinatorial background.  We compare the distribution of 
$d_{xy}$ of the pion candidate in the region 5600 $<$ M($B_c$) $<$ 6190
MeV/$c^2$ (lower side band) to that in the region 6390 $<$ M($B_c$)
$<$ 7200 MeV/$c^2$ (upper side band), where the main contribution
should be combinatorial.

\begin{figure}
\hfil
\epsfysize=2.65in
\epsffile {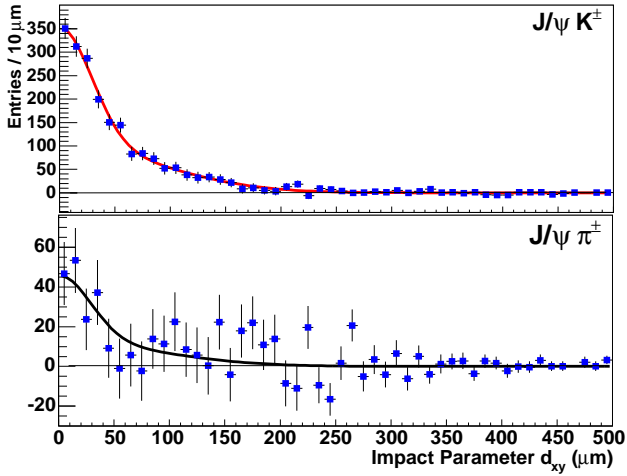}
\hfil
\label{fig:fdn}
\vspace*{-0.10in}
\caption{ 
Impact parameter 
of the third track relative to the $J/\psi$ vertex for the lower sideband region, after 
subtraction of the same distribution for the upper side band: 
(top) the curve is the sum of two Gaussians, fitted to the $B^{\pm}$ data points; 
(bottom) the $B_c^{\pm}$ data points,  overlaid with the above curve,  rescaled.
In both cases the selection criteria were relaxed.
}
\end{figure}

Figure 3 (top) shows the difference between the lower (4900-5100 MeV/c$^2$) 
and upper (5400-5700 MeV/c$^2$)
sidebands for the $d_{xy}$ distribution in the $B^{\pm}$ data 
sample, with
a large excess of events visible at small $d_{xy}$ values.  Figure 3
(bottom) shows the corresponding plot obtained using the $B_c^{\pm}$
candidate sample.  An enhancement is visible with a shape compatible
with that seen in the $B^{\pm}$ sample.  The $B^{\pm}$ curve, rescaled to 
fit the $B_c^{\pm}$ data, provides a good description of this
distribution.  The excess of low $d_{xy}$
events in the $B_c^{\pm}$ sample is evaluated to be 
244 $\pm$ 59,
where the uncertainty is statistical only. This result is consistent with 
Monte Carlo
estimates based on the calculations of \cite{Kiselev_03}. This supports 
the hypothesis that the broad 
physical background below the signal peak, evident in Fig. 2,
is in fact associated with partially reconstructed $B_c^{\pm}$ decays.

In conclusion, we observe a peak in the $J/\psi \pi^{\pm}$ mass
spectrum at a mass of $6285.7\pm5.3 (\rm stat)\pm1.2 (\rm syst)$
MeV/$c^2$. This peak is consistent with a narrow, weakly decaying particle 
state and is interpreted as the first evidence for fully
reconstructed decays of the $B_c^{\pm}$ meson. 
The mass value 
has much improved precision over the results obtained in $B_c^{\pm}$ 
semileptonic decays \cite{CDFRunI_bc,D0RunII_bc}.
There is also good agreement with recent theoretical predictions for
the $B_c^{\pm}$ mass 
around 6300 MeV/$c^2$~\cite{Brambilla_2002, QWGYR, lattice_2004}.

\begin{acknowledgments}
We thank the Fermilab staff and the technical staffs of the participating 
institutions for their vital contributions. We also thank A.V. Berezhnoy,
C.H. Chang and X.G. Wu for making available their calculations of 
$B_c^{\pm}$ production spectra.
This work was supported by the U.S. Department of Energy and National Science 
Foundation, the Italian Istituto Nazionale di Fisica Nucleare, the Ministry of
 Education, Culture, Sports, Science and Technology of Japan, 
the Natural Sciences and Engineering Research Council of Canada, 
the National Science Council of the Republic of China, the Swiss National 
Science Foundation, the A.P. Sloan Foundation, the Bundesministerium f\"ur 
Bildung und Forschung, Germany, the Korean Science and Engineering Foundation 
and the Korean Research Foundation, the Particle Physics and Astronomy 
Research Council and the Royal Society, UK, the Russian Foundation for Basic 
Research, the Comision Interministerial de Ciencia y Tecnologia, Spain, 
in part by the European Community's Human Potential Programme under contract 
HPRN-CT-2002-00292, and the Academy of Finland.

\end{acknowledgments}


\end{document}